\documentclass[preprint, showpacs,prl, floatfix, amssymb]{revtex4}
\usepackage{graphicx}
\usepackage{dcolumn}
\usepackage{bm}

\newcommand{\beq}{\begin{eqnarray}}
\newcommand{\eeq}{\end{eqnarray}}
\begin{document}

\title{Control of trapped-ion quantum states with optical pulses}
\author{C. Rangan$^a$, A.M.Bloch$^b$, C. Monroe$^a$, and P.H. Bucksbaum$^a$}
\affiliation{$^a$FOCUS Center and Departments of Physics and
 $^b$Mathematics, The University of
Michigan}
\date{December 5, 2003}

\begin{abstract}
We present new results on the quantum control of systems with
infinitely large Hilbert spaces. A control-theoretic analysis of
the control of trapped ion quantum states via optical pulses is
performed. We demonstrate how resonant bichromatic fields can be
applied in two contrasting ways --- one that makes the system
completely uncontrollable, and the other that makes the system
controllable. In some interesting cases, the Hilbert space of the
qubit-harmonic oscillator can be made finite, and the
Schr\"{o}dinger equation controllable via bichromatic resonant
pulses.  Extending this analysis to the quantum states of two
ions, a new scheme for producing entangled qubits is discovered.
\end{abstract}


\maketitle

Quantum computers rely on the quantum
coherence exhibited by physical systems, and on the understanding
of their control.  In this paper, we apply theoretical concepts of
quantum control to a scalable quantum-computing paradigm --- a
crystal of trapped ions~\cite{WinelandNISTreport1998}.  The two-level atom
(qubit) coupled to a harmonic oscillator is an example of
a quantum system with an infinitely large number of accessible
eigenstates.  The control of such systems is not well-understood.
Recently, it was argued based on compactness arguments
that infinite-dimensional systems are not
controllable~\cite{TuriniciReview2000}.  We provide a counter-example to
this long-held view~\cite{HuangJMP1983} --- the bichromatic control of
trapped-ion quantum states.

This work was motivated by the need to develop fast control
schemes to produce entangled states of qubits. Such entangled
states could then lead to interesting quantum states of the
coupled spin-motion system. There has been much interest recently
in applying techniques from coherent control to discover schemes
to perform quantum algorithms and universal gate
operations~\cite{RanganPRA2001}.  A recent paper shows a method of
creating fast control-phase gates in trapped
ions~\cite{GarciaRipollPRL2003}. We perform a theoretical analysis
of the control of trapped-ion quantum states with bichromatic
fields. We show that in some interesting cases of resonant
bichromatic control, the Hilbert space of the qubit-harmonic
oscillator system can be made finite. In these cases, the
Schr\"{o}dinger equation is controllable.  We show that the
controllability can be extended to two-ion quantum states, thus
providing a new scheme for qubit entanglement.

A trapped-ion qubit is most readily formed of two hyperfine states
of a laser-coolable ion, separated by a frequency $\omega_0/2\pi$
in the several GHz range. Qubits are coupled via the vibrational
modes of the ions' motion, which can be treated as quantum
harmonic oscillators~\cite{WinelandNISTreport1998}. The quantized
vibrational energy levels separated by a frequency $\omega_m/2\pi$
in the MHz range create sidebands in the spectrum of the ion. The
hyperfine `qubit' states are addressed by a pair of optical beams
(with counterpropagating wavevector components) in the Raman
(lambda) configuration, far detuned from an excited state that can
be adiabatically eliminated.  This interaction can be described by
a control Hamiltonian for an ion interacting with an
electromagnetic field $\bf{E}(\bf{\hat{r}},t) = E(t) \bf{\hat{x}}
\cos ( k z- \omega_L t)$ via an effective dipole moment ${\bf
\hat{\mu}}$~\cite{WinelandNISTreport1998}, and is written (in
atomic units, where $e\ = \ m_e\ =\ \hbar\ =\ 1$) as $  H_I  =
-\hat{\bf{\mu}}\cdot\bf{E(\bf{\hat{r}},t)}$. Choosing an
interaction picture where we can rotate away the fast $\omega_0$
contributions ($|\phi \rangle = \exp(\imath \frac{1}{2}\sigma_z
\omega_0t)|\psi\rangle$), and making the standard rotating wave
approximation, the field-free (drift) Hamiltonian $H_{0}^{\prime}
= \sum_m \omega_m \hat{n_m}$, and the control Hamiltonian is
written as
\begin{eqnarray}
  H_I^{\prime} = \sum_i\Omega^i(t)
  \left[\sigma^i_+e^{\imath\left[\Delta t+\sum_m M_m^i \eta_m(a_m
+a_m^{\dag})\right]}+h.c.\right]. \label{eq:Hint}
\end{eqnarray}
\noindent The Pauli operator ${\bf \sigma}$ describes the
equivalent spin-$\frac{1}{2}$ system represented by the qubit. The
detuning of the field's central frequency $\omega_L$ from
$\omega_0$, $\Delta$, is usually zero. The ion's position operator
$\hat{z}$ is expanded in normal mode coordinates as
$\hat{z}=z_{0m}(a_m +a_m^{\dag})$, $z_{0m}$ is the spatial extent
of the ground state wavefunction for the ion and mode being
addressed, $M_m^i$ is the relation (a matrix) between the position
of the $i^{th}$ ion and the normal mode coordinates. The
Lamb-Dicke parameter $\eta_m=kz_{0,m}$ is a coupling parameter
between an ion qubit and a motional mode when radiation at
wavevector $k$ is applied. $\Omega^i(t)=\frac{\mu^i E(t)}{4}$ has
the units of a Rabi frequency.

Firstly we present an analysis of the resonant control of the
coupled harmonic oscillator-spin-$\frac{1}{2}$ system in the
Lamb-Dicke limit (LDL), where the extent of the ions' motion is
much smaller than the wavelength of the applied field.  For
compactness, the superscript $i$ is dropped when discussing a
single ion.  We represent the various eigenstates
$|\hat{S},n_m\rangle$ by vertices of a graph as shown in Fig.1.
When a resonant electromagnetic field is applied, the coupling
between two eigenstates caused by the interaction form the edges.
This graph is similar to the `connectivity graph' of
Ref.~\cite{TuriniciReview2000}, and figures in
Ref.~\cite{LawOptExp1998}; with two differences that aid in our
understanding of the control mechanisms: the eigenstates are
ordered in energy, and the edges on the graph will represent the
matrix elements of the interaction between the eigenstates (not a
population flow between them), their thickness qualitatively
indicating the strength of the coupling.
\begin{figure}[h!]
\label{fig1} \centering
\rotatebox{0}{\includegraphics[width=2.75in]{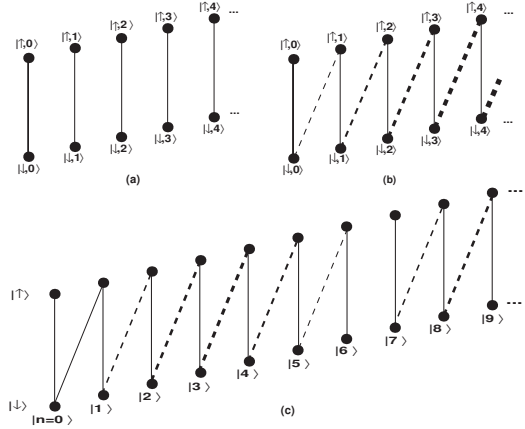}}
\caption{a: Lamb-Dicke limit (LDL), monochromatic field of
frequency $\omega_c$ , b: LDL, bichromatic field with frequencies
$\omega_c$ (solid edges) and $\omega_b$ (dashed edges), c: beyond
LDL, bichromatic field with frequencies $\omega_c$ and $\omega_b$.
The Lamb-Dicke parameter $\eta$ has been chosen so that the
$|\downarrow,6\rangle$ to $|\uparrow,7\rangle$ coupling vanishes.}
\end{figure}

For example, a field of frequency $\omega_L=\omega_c=\omega_0$
(the carrier frequency) acting on an ion connects states
$|\downarrow\ n\rangle$ and $|\uparrow\ n\rangle$. Similarly, a
field of frequency $\omega_L=\omega_{b}=\omega_0 + \omega_m$ (the
first blue sideband) connects states $|\downarrow\ n\rangle$ and
$|\uparrow\ n + 1\rangle$. In the LDL, it is well known that the
strength of the coupling due to the carrier is independent of the
phonon number of the eigenstates, whereas the couplings due to the
blue sideband increases as the square root of the vibrational
quantum number of the eigenstates.

We consider the simultaneous application of light of two
frequencies $\omega_c \  \& \ \omega_b$ in the LDL. The graph in
Fig.1(b) indicates that this bichromatic field makes the graph
transitively connected. That is, there exists a non-zero (direct
or indirect) coupling between every pair of eigenstates. But the
unbounded nature of the coupling operator leads to the
infinite-dimensionality of the accessible state space. The problem
is exacerbated because the strength of the coupling due to the
blue light increases as the square root of the vibrational quantum
number $n_m$. These features make the system uncontrollable, i.e.,
an arbitrary superposition of eigenstates cannot be created. In
fact, even a single eigenstate cannot be constructed. (For this
same reason, an infinite-level quantum harmonic oscillator is not
controllable with a resonant dipole field.)  This can be
rigorously demonstrated by calculating the Lie algebra formed by
the time-independent parts of Hamiltonians $H^{\prime}_0$ and
$H^{\prime}_I$, which does not close~\cite{Brockett1972}, and
whose span increases with successive elements of the algebra, thus
making it impossible to access only the desired state.

In many cases (such as Rydberg atoms or molecules), an
infinite-dimensional system can be numerically approximated by
truncating the state space to an essential (finite) basis.  This
is possible because the strength of the couplings due to the
applied field drops off outside this essential state space, and to
a good approximation, the coupling operator is bounded. In the
coupled spin-$\frac{1}{2}$-harmonic oscillator system, the
situation is exactly the opposite because of the infinitely many
degenerate transitions. Thinking in terms of control via
interfering paths~\cite{ShapiroJCP1986}, there are an infinite
number of likely paths that connect an initial state to a final
state.  Therefore truncating the state-space cannot approximate
this system.  This analysis indicates that there is no field made
of the carrier and blue frequencies that can completely control
the coupled harmonic oscillator-spin-$\frac{1}{2}$ system.  The
analysis is identical for bichromatic fields with the carrier and
red frequencies.

Contrary to this analysis, in 1996, Law and
Eberly~\cite{LawPRL1996} showed that by using the carrier and red
fields {\it alternately} it is possible to create arbitrary finite
superpositions of harmonic oscillator states.
Ref.~\cite{KneerPRA1998} shows that the same scheme can be used to
generate any finite superposition of states in a
spin-$\frac{1}{2}$/harmonic oscillator system.  This type of
control was also the basis of the quantum computing scheme
prescribed by Cirac and Zoller~\cite{CiracPRL1995}.  We now
analyze this scheme from the control theoretic viewpoint.

As seen in the graph in Fig.1(a), a monochromatic field reduces
the state space of the infinite-dimensional problem to (infinite
copies of) the two-level problem, and the resonant control of a
two-level system is well-understood. Indeed, using two colors
$\omega_0\ \&\ \omega_b$ alternately, any eigenstate can be
reached from any other eigenstate.  We can also interpret the
physics of this system by looking at the infinitesimal propagator
due to the bichromatic field, which is expanded using the
Baker-Campbell-Hausdorff formula.
\begin{eqnarray}
e^{i(H_c+H_b)\delta t)} & = & e^{i H_c\delta t}e^{i H_b\delta
t}e^{-\frac{1}{2}[H_c,H_b]\delta t^2}\\ \nonumber
                &   &
e^{-\frac{i}{12}([H_c,[H_c,H_b]]-[H_b,[H_c,H_b]])\delta
t^3}\cdots.
\end{eqnarray}
The higher the order of the commutator, the greater is the span of
the space produced by the propagator acting on an eigenstate.
Because the coupling strength of the blue transition increases
with increase in $n$, this series never truly truncates.  However,
if either $\Omega_c$ or $\Omega_b$ is turned off, the series
collapses beautifully to only the first or second term.  In this
way, a theoretically uncontrollable system can be made completely
controllable.  We note that in practice, the two colors should be
turned on adiabatically so that a single color field acts on the
system at any time.  This is a result contrary to the accepted
notion of control that the controllability of quantum systems does
not critically depend on the specific temporal profile of the
control field.

Thus, the Law-Eberly scheme~\cite{LawPRL1996} is the only explicit
scheme for accessing finite superpositions of trapped-ion quantum
states in the Lamb-Dicke limit. However, the method requires
sequential monochromatic pulses, each turned on and off
adiabatically. Therefore, it takes several trap periods to
complete a logic gate operation.  This motivated our investigation
of the control of the coupled spin-$\frac{1}{2}$/harmonic
oscillator system via shorter optical pulses with multiple
resonant colors.

In some interesting circumstances (described below), it is
possible to achieve bichromatic control over the system.  Consider a case when
the Lamb-Dicke criterion is not satisfied. By setting
$\omega_L=\omega_0$ in Eq.~\ref{eq:Hint}, a matrix element of the
interaction Hamiltonian in the field-free eigenbasis can be
written as
\begin{eqnarray}
\langle S^{\prime} n^{\prime}|H_I^{\prime}|S n \rangle \ = &
\Omega(t)2 {\rm Re}\left[\langle
S^{\prime}|\sigma_+|S\rangle \right.\quad \quad \quad \quad \quad \quad \quad \nonumber \\
 \otimes & \left.\langle n^{\prime}|\exp(\imath(M_m^i\eta_m(a_m
+a_m^{\dag })))|n \rangle\right],
\end{eqnarray}
\noindent where $M_m^i=1$ for one ion.  The harmonic oscillator
part of this matrix element~\cite{WinelandNISTreport1998} is
written as
\begin{eqnarray}
|\langle n^{\prime}|\exp(\imath(\eta_m(a_m +a_m^{\dag})))|n
\rangle|
\  = & \quad \nonumber \\
\quad \quad \quad \quad \quad \exp(-\eta_m
  /2)\sqrt{\frac{n_<!}{n_>!}}\  \eta_m^{|n^{\prime}-n|}
  & L_{n<}^{|n^{\prime}-n|}(\eta_m^2).\quad
\end{eqnarray}
\noindent The symbol $n_>$ refers to the larger of $n$ and
$n^{\prime}$, and $n_<$ refers to the smaller of $n$ and
$n^{\prime}$.  $L_n^{\alpha}(x)$ is the associated Laguerre
polynomial.  When the applied field contains the frequencies
$\omega_c$ and $\omega_b$, the harmonic oscillator parts of the
carrier and blue transition matrix elements have oscillatory
behaviors as shown in Fig.2. If the ion trap is adjusted (thereby
adjusting $\eta_m$) so that the coupling strength of one of the
(blue or carrier) transitions becomes zero, the system is
transformed into a finite closed subsystem, and a remaining
infinite subsystem. This is closely related to the technique of
truncating classical polynomials used to calculate numerical and
analytic solutions for the coherent dynamics of multi-level atoms
and molecules~\cite{Shore1977}. In Fig.1(c), the argument of the
Laguerre polynomial $\eta_m^2$ is adjusted to $0.527667$ so that
$L_6^1(\eta_m^2)=0$, and the $|\downarrow,6\rangle$ to
$|\uparrow,7\rangle$ transition is turned off. Similarly, the
$|\downarrow,5\rangle$ to $|\uparrow,5\rangle$ transition can be
turned off by adjusting $\eta_m^2 = 0.322548$ so that
$L_5^0(\eta_m^2)=0$. Experimentally, the Lamb-Dicke parameter can
be manipulated by adjusting the trap strength as shown by the
recent implementation of the wave packet CNOT
gate~\cite{DeMarcoPRL2002}.
\begin{figure}[h!]
\label{fig2} \centering
\rotatebox{0}{\includegraphics[width=2.25in]{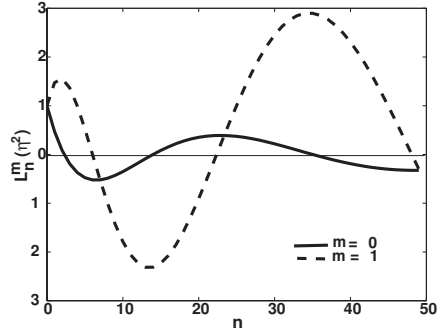}}
\caption{Associated Laguerre polynomial for the $|n\rangle
\rightarrow |n\pm m\rangle$ matrix elements, where $m=0$ for the
carrier transition, and $m=1$ for the blue sideband transition.}
\end{figure}

We now examine this finite harmonic oscillator/spin-$\frac{1}{2}$
system for controllability.  When the carrier and blue colors are
applied, it is seen in Fig.1(c) that the graph of this finite
subsystem is transitively connected.  Within the
completely-connected subspace, the fields cause degenerate
transitions.  However, the coupling strengths of the degenerate
transitions are unequal. Therefore, the conditions for
controllability of finite-dimensional
systems~\cite{Brockett1972,TuriniciReview2000} are satisfied. This
implies that there exists a target time $t_f<\infty$, when any
desired unitary transformation $U(0,t_f)$ can be performed within
this truncated Hilbert space using a bichromatic field.  Thus, by
using very few parameters: the two field strengths, the relative
phase between the two fields, and the target time, an arbitrary
superposition of eigenstates within the finite Hilbert space can
be created.

Note that this bichromatic control is still a relatively slow
process. However, there is no requirement of multiple adiabatic
turn-ons and turn-offs of the control fields, as in the
sequential, monochromatic control scheme.  It is expected that the
precision of the pulse timing in this scheme would have to be much
better than the time scale set by $\Omega^{-1}$, but this is also
true for conventional sequential schemes.  Also, the termination
of these resonant transitions and the faithful quantum control of
the system requires that the timescale of the control pulses be
much faster than the motional heating rate, which can cause the
motional quantum state to vacate the closed subsystem.  Motional
heating in ion traps has indeed been observed in all ion traps to
date~\cite{TurchettePRA2000}, but is not a fundamental limitation.
Recent experiments~\cite{RoweQIC2002} have observed motional
heating at much slower time scales than the control pulses
proposed here.

We now show how this bichromatic control scheme can be used to
produce entangled states of two qubits. For two trapped ions,
there are two axial modes of vibration - the center of mass mode,
and the breathing mode. Isolating one of the modes of vibration,
we consider the problem of producing an entangled state of spin
alone, that is, the vibrational contributions factor out.

Fig.~3 shows the graph of the two-ion single-mode eigenstates.  To
aid visibility, the hyperfine splittings $\omega_0^i$ of the two
ions is shown to be different.   For two ions of the same mass,
the magnitude of vibration is the same for both ions, therefore
$(M_m^i \eta_m)^2$ is the same. Therefore the trap can be adjusted
so that the spin-$\frac{1}{2}$ harmonic oscillator ladders
truncate at exactly the same transition for both ions. (This
truncation method does not work for more than two ions, because
then the magnitudes of vibration of all ions are not equal.) When
the two ions are individually addressed with bichromatic fields,
the relevant colors are $\omega_b^1,\ \omega_c^1,\ \omega_b^2$,
and $\omega_c^2$.  As seen in Fig.~3, the graph of the finite
subspace is transitively connected when fields $\omega_b^1,
\omega_b^2, \omega_c^1$ are applied. This graph satisfies all the
conditions required for the complete controllability of a finite
system~\cite{Brockett1972,TuriniciReview2000}.   If the hyperfine
splittings of the two ions are the same, $\omega_{b,c}^1$ is equal
to $\omega_{b,c}^2$.  This is a problem for controllability of the
finite system, which can be resolved by individually addressing of
the two ions.  This makes the degenerate transitions
distinguishable, and the two-ion single-mode system completely
controllable.  Thus, by individually addressing two ions with
resonant, bichromatic fields, it is possible to create an
arbitrary final state from any initially pure quantum state of
spin$+$motional finite subsystem.  In particular, entangled states
of two ion qubits can be created where the motion factors from the
overall quantum state.  For a scalable ion-trap quantum
computer~\cite{MonroeNature2002}, it is sufficient to generate
pairwise entanglement of two ion qubits.

\begin{figure}[ht!]
\label{fig3} \centering
\rotatebox{0}{\includegraphics[width=2.5in]{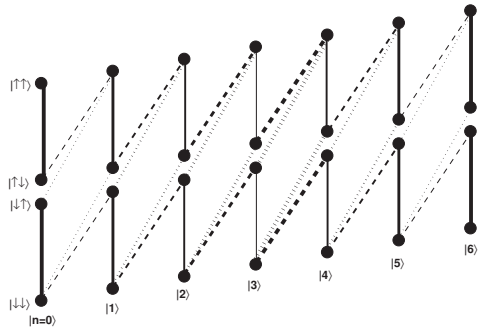}}
\caption{Finite subspace of two ions.  Three color resonant
field with $\omega_b^{1}$ (dashed line), $\omega_b^2$ (dotted line) and $\omega_c^{1}$(solid line). $\eta$ is adjusted
to create a finite subspace. }
\end{figure}

In contrast to the case of individual addressing, for multicolor,
resonant control with uniform illumination of the ions, it is
necessary to have ions with different $\omega_0$ for
controllability. This can be achieved experimentally by applying a
magnetic field gradient along the trap axis.  The gradient must be
judiciously chosen to avoid resonances caused by a coupling
between the two ions, or the r.f. field in the frame of the moving
ions, or the interference of the other motional mode levels.  More
practically, the gradient scheme sets a speed limit roughly given
by the differential Zeeman shift of the two ions. From our
controllability analysis, we know that entangled spin states can
be created using resonant, multicolor pulses. However, this
feature of having different $\omega_0$ makes it impractical to
calculate pulse solutions.  The interaction Hamiltonian is similar
to Eq.~\ref{eq:Hint}, with a summation over the terms for each
ion. If the $\omega_0^i$s are different, the detunings
$\omega_0^i-\omega_L$ cannot all be set to zero. The
time-dependent term in Eq.~\ref{eq:Hint} will make it necessary to
diagonalize the interaction matrix  at every time-step. This
increases the time to perform such calculations beyond reasonable
limits. Therefore, we propose an experiment where the three-color
pulse (with colors $\omega_b^1, \omega_b^2, \omega_c^1$) that can
produce entangled spin states of the trapped ions is generated via
learning control~\cite{GA}.  The amplitudes and phases of the
constituent colors form a small parameter space to be searched. It
will be expedient to experimentally implement the control of
trapped ions via three-color pulse shaping and allow the ions
themselves to `solve' the time-dependent Schr\"{o}dinger equation.
This scheme is an example of a quantum simulation of a calculation
that is intractable on a classical computer~\cite{Feynman1982}.

In summary, we have demonstrated that infinite-dimensional quantum
systems can be made completely controllable in certain
circumstances.  Our results show that the specific temporal shapes
of control fields are important in establishing controllability of
quantum systems. A bichromatic control scheme that leads to the
finite-dimensionality and controllability of the trapped-ion
system is presented.  This scheme can be extended to produce
entangled states of two trapped-ion qubits.

C.R. thanks the
NSF FOCUS Center for support, and the hospitality of
ITAMP (Harvard-Smithsonian Center for Astrophysics).  A.M.B. thanks the National Science Foundation for support.  We gratefully acknowledge helpful
conversations with Roger Brockett and Navin Khaneja at
Harvard University.

\end{document}